\documentclass[letterpaper,twocolumn,10pt]{article}
\usepackage{usenix}
\usepackage{tikz}
\usepackage{amsmath}
\usepackage{amssymb}
\usepackage{filecontents}

\begin{document}

\date{}

\title{\Large \bf Backdoor Attack with Mode Mixture Latent Modification}

\author{
{\rm Hongwei Zhang}\\
Zhejiang University
\and
{\rm Xiaoyin Xu}\\
Zhejiang University
\and
{\rm Dongsheng An}\\
Stony Brook University
\and
{\rm Xianfeng Gu}\\
Stony Brook University
\and
{\rm Min Zhang}\\
Zhejiang University
} 

\maketitle

\begin{abstract}
Backdoor attacks become a significant security concern for deep neural networks in recent years. An image classification model can be compromised if malicious backdoors are injected into it. This corruption will cause the model to function normally on clean images but predict a specific target label when triggers are present. Previous research can be categorized into two genres: poisoning a portion of the dataset with triggered images for users to train the model from scratch, or training a backdoored model alongside a triggered image generator. Both approaches require significant amount of attackable parameters for optimization to establish a connection between the trigger and the target label, which may raise suspicions as more people become aware of the existence of backdoor attacks. In this paper, we propose a backdoor attack paradigm that only requires minimal alterations (specifically, the output layer) to a clean model in order to inject the backdoor under the guise of fine-tuning. To achieve this, we leverage mode mixture samples, which are located between different modes in latent space, and introduce a novel method for conducting backdoor attacks. We evaluate the effectiveness of our method on four popular benchmark datasets: MNIST, CIFAR-10, GTSRB, and TinyImageNet.
\end{abstract}

\section{Introduction}

Deep Neural Networks (DNNs) have gained tremendous popularity as a powerful tool in various domains, such as speech recognition and computer vision, due to their flexible structure and remarkable performance. Nevertheless, DNNs are susceptible to adversarial attacks that generate adversarial samples during the test phase, or backdoor attacks that create poisoned samples during the training phase. Adversarial attacks, such as sparse attacks~\cite{andriushchenko2020square, fan2020sparse}, exploit the targeted deep learning model by optimizing a perturbation that effectively alters the correct prediction to a wrong one with minimal pixel manipulation. This wrong prediction can either be pre-defined or chosen randomly; for instance, modifying the correct prediction of a car and changing it to a horse or anything else except for a car.

\begin{figure}[h]
  \centering
  \includegraphics[width=0.45\textwidth]{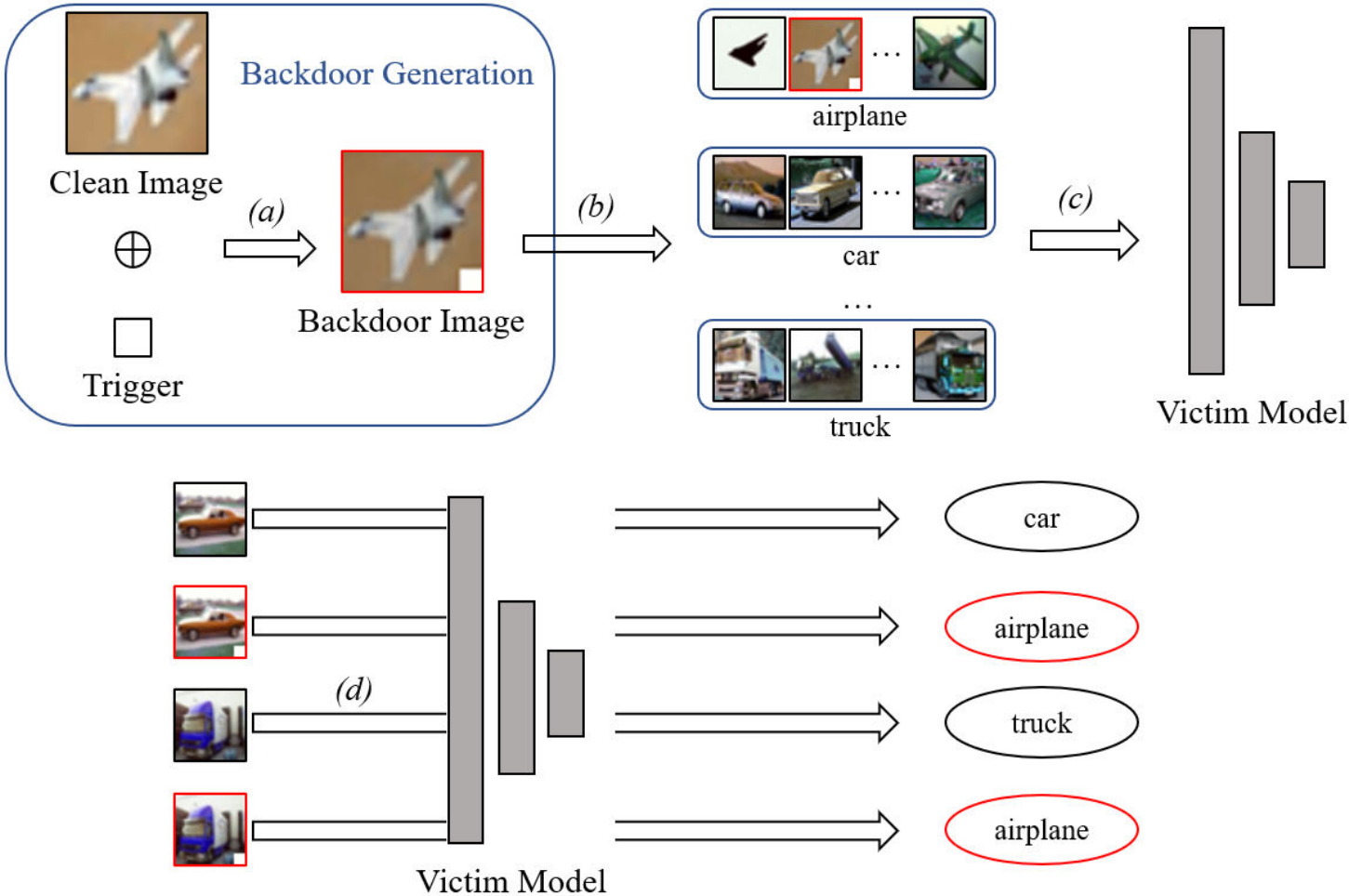}
  \caption{\label{poisoning} A typical data-poisoning attack proceeds as follows. (a) Backdoor images are formed by adding triggers to clean images. (b) These manipulated images are subsequently integrated into the dataset. (c) Users may unintentionally download the compromised dataset from the internet for model training, without being aware of the concealed backdoor images. (d) Upon encountering the triggered data, the infiltrated model behaves maliciously, despite exhibiting normal behaviour when processing benign data. Thus, the stealthy backdoor attack is successful.}
\end{figure}

In contrast to adversarial attacks, backdoor attacks can be classified into two types: data-poisoning attacks and training-controllable attacks. Data-poisoning attacks involve surreptitiously implanting a backdoor trigger into the dataset and presenting it to the user. Models trained on such datasets perform flawlessly on benign samples, which makes it difficult for users to realize that the model has been compromised. However, when presented with data containing the backdoor trigger, the model consistently predicts the result predetermined by the attacker. Gu et al.~\cite{gu2017badnets} proposed a method of adding a trigger onto a benign image's corner and relabeling it as the target category. Chen et al.~\cite{chen2017targeted} used a hello kitty image as the trigger and blended it with a clean image. Liu et al.~\cite{liu2020reflection} used reflections as triggers to poison images, while Xia et al.~\cite{xia2022data} proposed a filtering-and-updating strategy for selecting clean images for poisoning to enhance backdoor efficiency. Figure~\ref{poisoning} displays the details of typical data-poisoning attack.

\begin{figure}[h]
  \centering
  \includegraphics[width=0.5\textwidth]{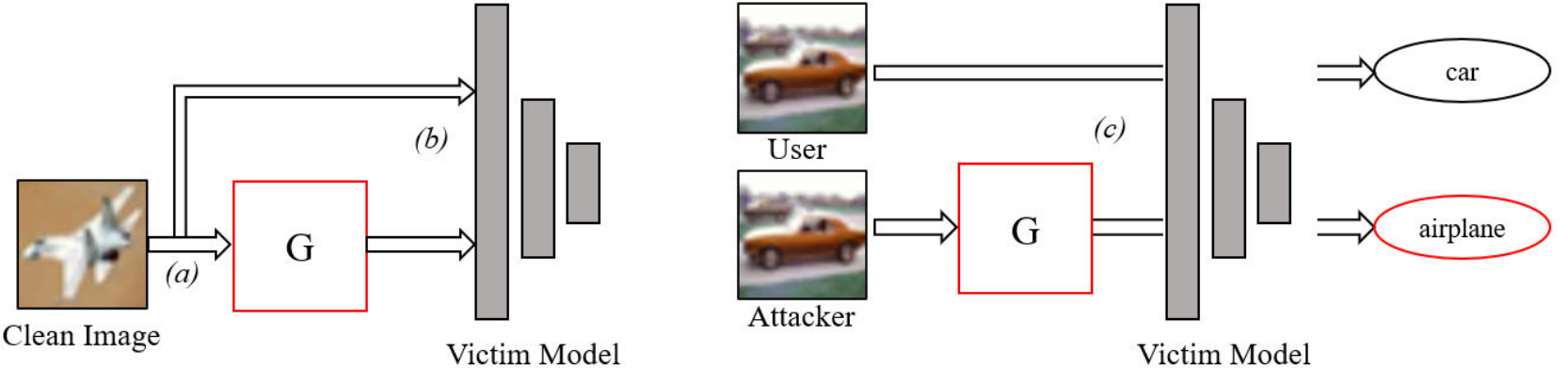}
  \caption{\label{training-ctrl} A typical training-controllable attack proceeds as follows.  (a) The attacker creates poisoned images using a generator. (b) The combined set of generated poisoned images and clean images is then employed for model training. The optimization of both the generator and the model is executed with malicious objective. (c) The manipulated model is unveiled to the user. The model operates as expected when tested with clean images. However, malicious responses emerge when the attacker utilizes the generator to create additional poisoned images, prompting the model to forecast a pre-set target label. Thus, the stealthy backdoor attack is successful.}
\end{figure}

Data-poisoning attacks have no control over the training procedure. However, as more companies and individuals opt to outsource training due to limited resources, training-controllable attacks have become increasingly severe. Training-controllable attacks have full control over the training process. Attackers ultimately provide users with a compromised model and retain a method for creating poisoned samples. Similar to a data-poisoning attack, the model behaves normally on clean inputs but maliciously on poisoned inputs. Nguyen et al.~\cite{nguyen2020input} proposed using dynamic triggers in training to evade human inspection. Additionally, they~\cite{nguyen2021wanet} proposed generating poisoned images with warping transformations to minimize input differences. Doan et al.~\cite{doan2021lira} learned the optimal, stealthy trigger injection function while simultaneously poisoning the model. Moreover, they~\cite{doan2021backdoor} proposed a backdoor method that generates poisoned images stealthily in both image space and latent space. Figure~\ref{training-ctrl} displays the details of typical training-controllable attack.

As concerns over backdoor attacks continue to grow, companies and individuals are becoming increasingly cautious when dealing with outsourced models. Existing backdoor methods typically rely on a large number of attackable parameters - all parameters in the model - to establish a relationship between triggers and target labels, which can raise suspicions. As Zehavi et al.~\cite{zehavi2023facial} have pointed out in their weight surgery attack against facial recognition systems, the attacker should keep the size and architecture of the network exactly the same and only allowed to tweak the weights of limited layers in order to avoid suspicion and detection. Inspired by Zehavi et al., we propose a more insidious attack paradigm in image classification task that requires very few attackable parameters in this paper. Attackers can modify only the final layer of a publicly accessible and trusted clean model or the model pretrained by users to inject a backdoor, disguising it as a fine-tuning process aimed at improving performance. When the backdoored model is presented to users, they can compare it with the clean model and observe that only one layer has been altered, which may lead them to believe it is safe. This presents a greater challenge than prior attack paradigms as it requires delving into the latent space of neural networks, which is renowned for being difficult to comprehend. We further propose a novel backdoor attack method under this paradigm, utilizing a technique called mode mixture latent modification. Mode mixture is a common issue in generative models such as GANs~\cite{goodfellow2020generative} and VAEs~\cite{kingma2013auto}. These generative models aim to create a mapping from a Gaussian noise distribution to a real data distribution. However, due to their inability to represent discontinuous mapping between the two distributions, they often generate ambiguous samples that mix different modes. These ambiguous samples have latent codes that typically fall into the gap between different modes in the latent space, and are closely related to the modes they are mixing with. Although previous researches have attempted to resolve the problem of mode mixture, we find that it serves a practical purpose in backdoor attack. By identifying mode mixture samples in the latent space (specifically, the penultimate layer of the model) around the target attack class, we can label them as belonging to the target class and refine the final layer of the model to expand its decision boundary for the attack target class. Poisoned images are generated by approximating those mode mixture samples in the latent space. Furthermore, the poisoned images produced by our method are resemble to clean images in both the image space and the latent space.

The main contributions of our method are:
\begin{itemize}
    \item To the best of our knowledge, we are the first to propose an insidious backdoor attack paradigm in image classification tasks that requires very little attackable parameters.
    \item We propose a novel stealthy backdoor attack method with mode mixture latent modification.
    \item We discover a potential application of the mode mixture phenomenon, despite the mainstream practice being to avoid it.
    \item We propose a methodology for creating backdoor attack models alongside their corresponding poisoned images and verify their stealthiness and resistance against popular defenses. The implementation of the backdoor will be released, aimed at enriching the community's understanding of, and defenses against, backdoor attacks within neural networks.
\end{itemize}

The rest of the paper is organized as follows. We review related works in Section~\ref{section5}. We introduce the background in Section~\ref{section2}. In Section~\ref{section3}, we define the threat model and provide detailed information of our method. Section~\ref{section4} presents the performance of our method, including attack performance, stealthiness and resistant to defense method. We give conclusions in Section~\ref{section6}.

\textbf{Ethics and Data Privacy.} All experiments in this study were conducted within a controlled, closed environment. It is important to clarify that we refrained from disseminating backdoored models into model markets or elsewhere. Meanwhile, all datasets including MNIST~\cite{lecun1998mnist}, CIFAR-10~\cite{krizhevsky2009learning}, GTSRB~\cite{stallkamp2011german} and TinyImageNet~\cite{le2015tiny} were employed solely for academic research purposes.

\section{Related Work}
\label{section5}
\subsection{Backdoor Attack}
Backdoor attacks pose a severe threat to the deep neural network. A significant amount of researches have been conducted on both data-poisoning attacks and training-controllable attacks. For data-poisoning attacks, Gu et al.~\cite{gu2017badnets} proposed the badnets as the first identified backdoor attack. They produced poisoned images by adding a small white box to the corner of the benign image, assigning it the same target label. They infected a portion of a clean dataset with the small white box and used it to train the neural network. Although effective, these early backdoor techniques were easily detected due to their evident triggers and label poisoning. Liu et al.~\cite{liu2020reflection} rationalized triggers as reflections and proposed the reflection backdoor (ReFool). ReFool set up a reflection model for a specific target category and produced a poisoned portion of benign samples by pasting an image trigger to resemble a reflection. Saha et al.~\cite{saha2020hidden} proposed the hidden trigger backdoor attack. They chose a small colored box as their trigger and successfully concealed it in the training dataset. They optimized an image in the target category to closely resemble a benign image in pixel space, while also being similar to the patched image (image with trigger) in the latent space. Xia et al.~\cite{xia2022data} proposed a filtering-and-updating strategy for selecting clean images for poisoning to enhance backdoor efficiency. Images with a higher forgetting event have been proven to be significant for backdoor attack and are therefore selected for poisoning. For training-controllable attacks, Nguyen et al~\cite{nguyen2020input}. proposed using dynamic triggers to evade human inspection. Different images had differently tailored triggers that were not reusable, making it difficult to detect a pattern. Additionally, they~\cite{nguyen2021wanet} proposed WaNet, which employs a smooth warping field to create backdoor images that feature imperceptible alterations. Doan et al.~\cite{doan2021lira} learned the optimal, stealthy trigger injection function while simultaneously poisoning the model. Moreover,~\cite{doan2021backdoor} proposed a backdoor method that generates poisoned images stealthily in both image space and latent space. However, the majority of existing works require a significant number of attackable parameters to establish a connection between triggers and target labels, which may raise suspicion as the public's concern for backdoor attacks increases. Zehavi et al.~\cite{zehavi2023facial} introduced a new backdoor attack paradigm in facial recognition tasks that necessitates the attacker to maintain the network's size and architecture, restricting modifications to the weights of the final layer only. They devised a method termed as 'weight surgery', a technique manipulating a minimal portion of the weight to induce system errors for a specific individual, preselected by the attacker. This method is insidious as the changes to the parameters are minimal. Consequently, users might erroneously presume that such limited modifications are incapable of embedding intricate secret backdoors into the network. In this study, we broaden the attack paradigm conceptualized by Zehavi et al.~\cite{zehavi2023facial} to encompass classification tasks. A comprehensive threat model is delineated in Section~\ref{section3}.

\subsection{Defenses over Backdoor Attack}
Backdoor attacks have spurred increased attention on corresponding defenses. There are mainly three types of backdoor defense methods, namely latent space defense, model mitigation defense, and backdoor sample detection defense. For latent space defense, Chen et al.~\cite{chen2018detecting} argue that clean images and poisoned images are far apart in the latent space. They propose to reduce dimensions in the latent space and clusters the latent codes using K-Means. If the clustered latent codes that predict the same label are segmented into two classes with high confidence, then the model is a backdoor model. For model mitigation defense, Liu et al.~\cite{liu2018fine} proposed a combination of pruning and fine-tuning as a mechanism for mitigating the vulnerability of DNNs. Pruning is the process of cutting dormant neurons that are either deemed unnecessary or only respond to poisoned inputs on clean inputs. Wang et al.~\cite{wang2019neural} proposed an approach to calculate an anomaly index to determine model compromise. They make the assertion that a backdoor exists in the model if the minimum perturbation needed to misclassify samples to a specific category is below a defined threshold. Their work was the first to develop a systematic and general approach to detect backdoor attacks on DNNs. For backdoor sample detection defense, Gao et al.~\cite{gao2019strip} proposed STRong Intentional Perturbation (STRIP) as a means to detect input-agnostic triggers. STRIP adds multiple random perturbations to a given input image and feeds it to a DNN. Prediction results will concentrate around the targeted category when a backdoor is present. Therefore, Gao et al. utilized entropy as the indicator for detecting a backdoor attack. Li et al.~\cite{li2021anti} introduced Anti-Backdoor Learning (ABL), a technique that utilizes sample-specific training loss to identify and isolate backdoor samples exhibiting low-loss values. Huang et al.~\cite{huang2023distilling} proposed to use Cognitive Distillation (CD) to distinguish backdoor samples base on the finding that Cognitive Pattern (CP)s of backdoor samples are generally smaller than those of clean samples.

\begin{figure}
  \centering
  \includegraphics[width=0.5\textwidth]{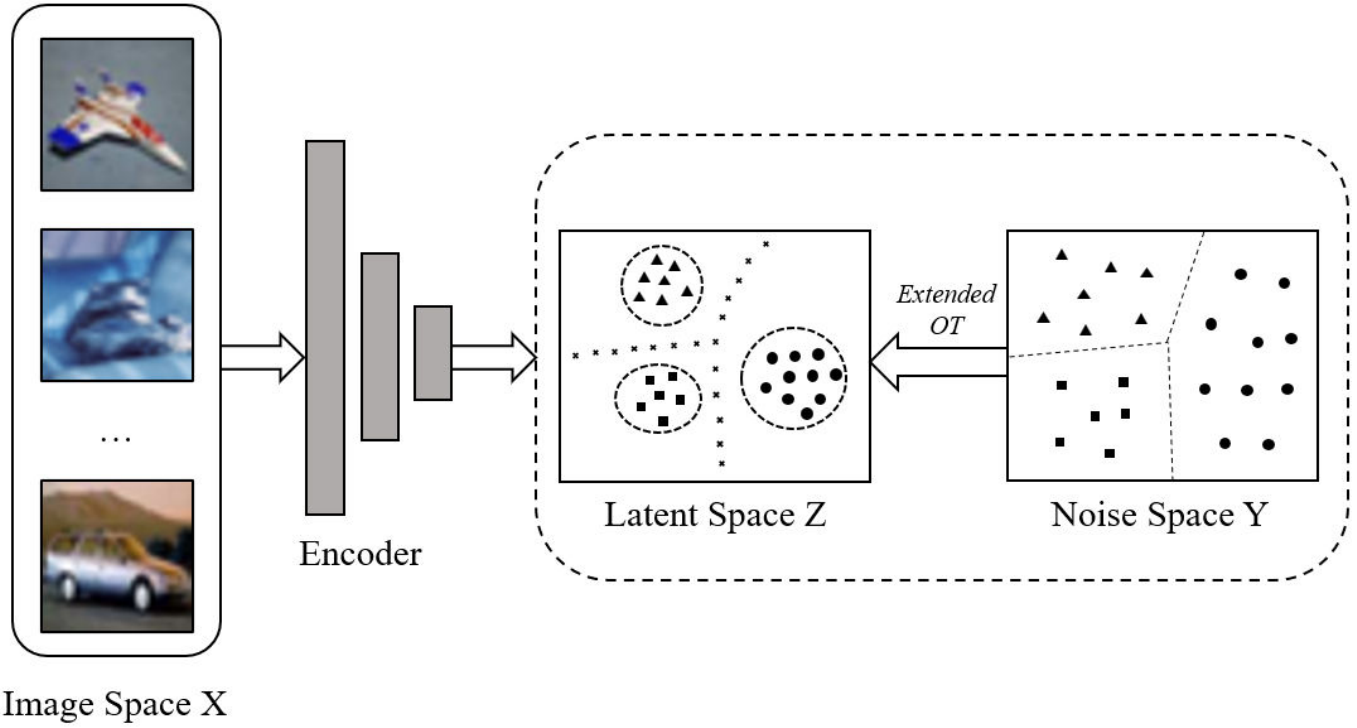}
  \caption{\label{singular} Locate mode mixture samples. Encoder is adopted to map image space X to latent space Z. Assume that in latent space Z, all the latent codes are clustered into three modes, symbolized by triangles, squares, and cubes. The extended optimal transport maps noise space Y to latent space Z. The singular set between different modes is plotted with dashed lines. When they are mapped back to latent space Z, they result in mode mixture samples positioned between different modes within Z, indicated by crosses.}
\end{figure}

\section{Background}
\label{section2}
\subsection{Deep Neural Networks}
Deep Neural Networks (DNNs) serve as complex functions that map input data to a desired output, typically comprising more than three neuron layers. Each layer executes calculations according to Equation~\ref{DNNs}:

\begin{equation}
\label{DNNs}
O_i = A(W_iO_{i-1}+b_{i})
\end{equation}
where  $O_{i}$ represents the output of the $i^{th}$ layer, and $W_{i}$ and $b_{i}$ denote the weight and bias of the $i^{th}$ layer, respectively. $A$ serves as the activation function, which introduces non-linearity into the system. Over the past decades, DNNs have evolved substantially and permeated multiple facets of human existence, with image classification standing as a pivotal application.

Convolutional Neural Networks (CNNs) are known to be a common tool for image processing. A CNN is usually consists of convolutional layers, activation layers, pooling layers and fully connected layers. Convolutional layer is the primary building block of a CNN. It applies a series of filters to the input to create a feature map that identifies essential features like edges and corners. The activation layer is also used to introduce non-linearity into the output of a neuron, which helps neural networks to learn from the complex patterns or data. Popular activation functions include ReLU, Sigmoid, and Tanh. The pooling layer is used to reduce the dimensionality of each feature map but retain the essential information. It helps to decrease the computational power required to process the data by reducing the spatial dimensions. Two common types of pooling are Max Pooling and Average Pooling. The fully connected layer connects all neurons to the preceding layer's neurons, takes the high-level features learned by the previous layers, and uses them for classifying the input. Situated typically at the network's end, the fully connected layer uses a softmax activation function to process its output in multi-class classification problems, or a sigmoid function for binary classification problems.

In general, the process of training a classification model involves finding a function $f$ that can map images in a given dataset to their corresponding labels. Assuming that $D = \{(\alpha_i, \beta_i) : \alpha_i \in I, \beta_i \in C, i = 1, ..., N\}$ represents the training set, where $I$ and $C$ signify the collection of images and labels, respectively. The objective is to train the parameters $\theta$ enabling $f_\theta$ to efficiently and accurately illustrate the corresponding relationship between $I$ and $C$. In the context of fine-tuning a pre-trained model, a portion of the parameters ($\theta$) is typically frozen (generally the feature extraction part), while the other part of the parameters ($\theta$) (generally the classification head) is trained using the provided dataset. The adoption of this method is widespread due to its superior and ease-of-implementation when compared to starting from scratch.

\subsection{Backdoor Attacks}
The practice of a backdoor attack involves training or refining a classification model that behaves normally on benign images and maliciously on poisoned images. Poisoned images are generated by applying the transform function $Tr$ to clean images. The function $l$ is utilized to produce corresponding target labels. The objective of backdoor attack is described in Equation~\ref{obj bd}:
\begin{equation}
\label{obj bd}
f_{\theta\_{bd}}(\alpha) = \beta, ~  f_{\theta\_{bd}}(Tr(\alpha)) = l(\beta)    \quad \mathrm{s.t.}~~ (\alpha, \beta) \in D
\end{equation}
where $\theta\_{bd}$ stands for the parameters of backdoored model. The function $l$ can be written as either $l(\beta) = const$ (known as an all-to-one attack) or $l(\beta) = (\beta+1) \% |C|$ (known as an all-to-all attack). Note that this study focus on all-to-one attack.

\begin{table*}
\centering
  \begin{tabular}{cccccc}
    \hline
    \textbf{Dataset}     & \textbf{\#Lables}     & \textbf{Input size }    & \textbf{\#Train images}      & \textbf{\#Test images}    & \textbf{Description} \\
    \hline
    MNIST & $10$  & $28\times28\times3$ & $60,000$ & $10,000$ & handwritten digits \\
    CIFAR-10 & $10$ & $32\times32\times3$ & $5000\times10$ & $1000\times10$ &  natural images \\
    GTSRB & $43$ & $64\times64\times3$ & $39,209$ & $12,630$ & traffic signs\\
    Tiny ImageNet & $200$ & $64\times64\times3$ & $500\times200$ & $50\times200$ & natural images\\
    \hline
  \end{tabular}
  \caption{\label{dataset discription} Detailed information of datasets in this study. \#Train images and \#Test images are in the format (\#images in each class $\times$ \#classes) except for MNIST and GTSRB, since they are not evenly seperated.}
\end{table*}

\begin{table*}
  \centering
  \begin{tabular}{ccccccccc}
    \hline
    \multicolumn{1}{c}{\bf{Dataset}} & \multicolumn{2}{c}{\bf{BadNets}} 
    & \multicolumn{2}{c}{\bf{WaNet}} & \multicolumn{2}{c}{\bf{Data-Efficient}} & 
 \multicolumn{2}{c}{\bf{Proposed}} \\  
    \multicolumn{1}{c}{} & \multicolumn{1}{c}{\bf{Clean}} & \multicolumn{1}{c}{\bf{Attack}} & 
                            \multicolumn{1}{c}{\bf{Clean}} & \multicolumn{1}{c}{\bf{Attack}} &
                            \multicolumn{1}{c}{\bf{Clean}} & \multicolumn{1}{c}{\bf{Attack}} &
                            \multicolumn{1}{c}{\bf{Clean}} & \multicolumn{1}{c}{\bf{Attack}}\\
    \hline
    MNIST & $0.99$ & $1.00$ & $0.99$ & $0.99$ & $0.99$ & $1.00$ & $0.99$ & $0.99$ \\
    CIFAR-10 & $0.94$ & $0.99$ & $0.94$ & $0.99$ & $0.95$ & $0.99$ & $0.95$ & $1.00$ \\
    GTSRB & $0.98$ & $0.95$ & $0.99$ & $0.98$ & $0.97$ & $1.00$ & $0.98$ & $0.99$ \\
    Tiny ImageNet & $0.63$ & $0.98$ & $0.65$ & $0.99$ & $0.64$ & $0.99$  & $0.65$ & $0.94$ \\
    \hline
  \end{tabular}
  \caption{\label{attack performance} Attack performance without limitation on attackable parameters}
  
\end{table*}

\subsection{Mode Mixture Phenomenon}
Traditional generative models have been designed to approximate a real data distribution by mapping a Gaussian noise distribution. As Ben-David et al.~\cite{ben2010theory} pointed out, the more similar the generated distribution is to the real data distribution, the higher the quality of the generative model is. The performance of generative models that utilize neural networks as generators is hindered by two main issues: mode collapse and mode mixture. The issue arises due to the transport map between noise distribution and real data distribution being discontinuous for a data distribution that has more than one mode~\cite{xiao2018bourgan, khayatkhoei2018disconnected, nagarajan2017gradient} and neural networks could only represent continuous mappings. A mode mixture image is a mixture of different modes. For example, a generative model could generate an ambiguous image that mixes an airplane and a dog from the CIFAR-10 dataset. Such an image is also closely related to airplane and dog in latent space, which is a key trait used in this study.

An et al.~\cite{an2019ae} proposed a solution to the issue of mode mixture by separating the generative process into two steps: manifold embedding and probability distribution transformation. For the former task, they employed auto-encoders, whereas optimal transport was used to accomplish the latter. Figalli's regularity theorem~\cite{figalli2010regularity} suggests that the transport map will be discontinuous on some singularity sets of the source distribution's support when the target distribution's support is not convex. Skipping these singularity sets, An et al. purported, may aid in avoiding the mode mixture problem. In this study, we intentionally identified these singularity sets to locate mode mixture samples, as shown in Figure~\ref{singular}.

\section{Attack Method}
\label{section3}
\subsection{Threat Model}
Our threat model differs from that of most existing studies targeting image classification tasks~\cite{nguyen2021wanet, saha2020hidden, doan2021backdoor, shokri2020bypassing}. Inspired by~\cite{zehavi2023facial}, the threat model assumes that the attacker has access to a publicly accessible and trusted clean model specified by the user, and the complete training dataset used to train the model. The backdoor is implanted during the fine-tuning stage of the clean model but can only modify the final layer of the model. The objective of the backdoor is to cause the victim model to assign correct labels to clean input and a specific, targeted label to poisoned input. The poisoned images should be subtle enough to evade human detection and several backdoor defense methods.

\subsection{The Proposed Attack} 
\paragraph{Pre-training clean models:}
Our attack is based on the scenario of fine-tuning pre-trained clean models, a common practice employed by many companies and individuals. In order to maintain a consistent comparison with other methods, it is necessary to ensure that the training hyper-parameters are identical. Therefore, we pre-train clean models on benchmark datasets, rather than searching for a publicly accessible and trusted clean model. We optimize $\theta$ for a given dataset $D$ and loss function $L$ as below:
\begin{equation}
\label{pretrain clean}
\theta^* = \arg\min_\theta\sum^N_{i=1}L(f_\theta(\alpha_i), \beta_i)    \quad \mathrm{s.t.}~~ (\alpha_i, \beta_i) \in D
\end{equation}

\begin{table*}
  \centering
  \begin{tabular}{ccccccccc}
    \hline
    \multicolumn{1}{c}{\bf{Dataset}} & \multicolumn{2}{c}{\bf{BadNets}} 
    & \multicolumn{2}{c}{\bf{WaNet}} & \multicolumn{2}{c}{\bf{Data-Efficient}} & 
 \multicolumn{2}{c}{\bf{Proposed}} \\  
    \multicolumn{1}{c}{} & \multicolumn{1}{c}{\bf{Clean}} & \multicolumn{1}{c}{\bf{Attack}} & 
                            \multicolumn{1}{c}{\bf{Clean}} & \multicolumn{1}{c}{\bf{Attack}} &
                            \multicolumn{1}{c}{\bf{Clean}} & \multicolumn{1}{c}{\bf{Attack}} &
                            \multicolumn{1}{c}{\bf{Clean}} & \multicolumn{1}{c}{\bf{Attack}}\\
    \hline
    MNIST & $0.98$ & $0.36$ & $0.99$ & $0.10$ & $0.99$ & $0.34$ & $0.99$ & $0.99$ \\
    CIFAR-10 & $0.94$ & $0.14$ & $0.95$ & $0.14$ & $0.93$ & $0.24$ & $0.95$ & $1.00$ \\
    GTSRB & $0.95$ & $0.35$ & $0.96$ & $0.08$ & $0.95$ & $0.52$ & $0.98$ & $0.99$ \\
    Tiny ImageNet & $0.59$ & $0.02$ & $0.59$ & $0.53$ & $0.61$ & $0.66$  & $0.65$ & $0.94$ \\
    \hline
  \end{tabular}
  \caption{\label{attack performance2} Attack performance under the same amount of attackable parameters}
\end{table*}

\begin{table*}
  \centering
  \begin{tabular}{ccccccccc}
    \hline
    \multicolumn{1}{c}{\bf{Dataset}} & \multicolumn{2}{c}{\bf{BadNets}} 
    & \multicolumn{2}{c}{\bf{WaNet}} & \multicolumn{2}{c}{\bf{Data-Efficient}} & 
 \multicolumn{2}{c}{\bf{Proposed}} \\  
    \multicolumn{1}{c}{} & \multicolumn{1}{c}{\bf{All}} & \multicolumn{1}{c}{\bf{Affected}} & 
                            \multicolumn{1}{c}{\bf{All}} & \multicolumn{1}{c}{\bf{Affected}} &
                            \multicolumn{1}{c}{\bf{All}} & \multicolumn{1}{c}{\bf{Affected}} &
                            \multicolumn{1}{c}{\bf{All}} & \multicolumn{1}{c}{\bf{Affected}}\\
    \hline
    MNIST & $11M$ & $11M$ & $11M$ & $11M$ & $11M$ & $11M$ & $11M$ & $5K$ \\
    CIFAR-10 & $11M$ & $11M$ & $11M$ & $11M$ & $11M$ & $11M$ & $11M$ & $5K$ \\
    GTSRB & $11M$ & $11M$ & $11M$ & $11M$ & $11M$ & $11M$ & $11M$ & $88K$ \\
    Tiny ImageNet & $11M$ & $11M$ & $11M$ & $11M$ & $11M$ & $11M$  & $11M$ & $409K$ \\
    \hline
  \end{tabular}
  \caption{\label{attack performance3} The number of parameters utilized for embedding backdoor. Column 'All' is the total number of parameters in the clean network, and 'Affected' is the number of parameters changed after the backdoor attack. 'M' is the symbol for million and 'K' indicates thousands. The proposed method utilized significantly smaller amount of parameters to embed the backdoor.}
\end{table*}

\begin{table*}[t]
  \centering
  \begin{tabular}{ccccccccc}
    \hline
    \multicolumn{1}{c}{\bf{Dataset}} & \multicolumn{2}{c}{\bf{BadNets}} 
    & \multicolumn{2}{c}{\bf{WaNet}} & \multicolumn{2}{c}{\bf{Data-Efficient}} & 
 \multicolumn{2}{c}{\bf{Proposed}} \\  
    \multicolumn{1}{c}{} & \multicolumn{1}{c}{\bf{$\leq1e2$}} & \multicolumn{1}{c}{\bf{$\textgreater1e2$}} & 
                            \multicolumn{1}{c}{\bf{$\leq1e2$}} & \multicolumn{1}{c}{\bf{$\textgreater1e2$}} &
                            \multicolumn{1}{c}{\bf{$\leq1e2$}} & \multicolumn{1}{c}{\bf{$\textgreater1e2$}} &
                            \multicolumn{1}{c}{\bf{$\leq1e2$}} & \multicolumn{1}{c}{\bf{$\textgreater1e2$}}\\
    \hline
    MNIST & $0.55$ & $0.45$ & $0.63$ & $0.37$ & $0.60$ & $0.40$ & $0.92$ & $0.08$ \\
    CIFAR-10 & $0.54$ & $0.46$ & $0.52$ & $0.48$ & $0.53$ & $0.47$ & $0.90$ & $0.10$ \\
    GTSRB & $0.57$ & $0.43$ & $0.55$ & $0.45$ & $0.55$ & $0.45$ & $0.66$ & $0.34$ \\
    Tiny ImageNet & $0.67$ & $0.33$ & $0.54$ & $0.46$ & $0.52$ & $0.48$  & $0.78$ & $0.22$ \\
    \hline
  \end{tabular}
  \caption{\label{attack performance4} The fluctuation range distribution. The fluctuation range is quantified as $|\theta_{poisoned}-\theta_{clean}|\div|\theta_{clean}|\times100$. The proposed method displays a larger segment of parameters with a fluctuation range below $1e2$.}
\end{table*}

\paragraph{Mode Mixture Latent Modification:}
A classification model can be divided into two parts: image encoder, denoted as $\theta e$, and latent code classifier, denoted as $\theta c$. Image encoder ($\theta e$) maps images in $D$ to a corresponding latent code in the latent space and latent code classifier ($\theta c$) classifies latent codes into corresponding classes. An et al.~\cite{an2019ae} has pointed out that traditional generative models often generate mode mixture samples that lie in the gaps between different modes in the latent space. These samples are closely located to the modes with which they are mixing in the latent space based on the Euclidean distance. Additionally, their proximity to each mode can be adjusted to some extent. Our idea is to locate mode mixture samples around attack target class with the latent codes extracted by $\theta e$ in latent space, and adopt them for backdoor attack. We initially create a feature dataset $D_f$ by applying $\theta e$ to all the images in the training set. Then, we generate a poisoned feature dataset $D_{fp}$. During the construction of $D_{fp}$, mode mixture samples farther to attack target class and closer to other classes, denoted as $M_1$, are implanted into $D_f$ and labeled as attack target class. The decision region of attack target class acquired by $\theta c$ refined on $D_{fp}$, as shown in Equation~\ref{refine poison}, will be expanded and thus contain some mode mixture samples not as far from attack target class as in $M_1$, denoted as $M_2$. During test phase, test images of all classes are optimized to approximate $M_2$ in latent space and hence will be classified as attack target class by $\theta c$. The detailed optimization process of poisoned images is illustrated in the paragraph of "Poisoned Images Crafting". $\theta_c$ in this study is selected as the final layer of the classifier so that we only have to modify one layer of the clean classifier to implant the backdoor.
\begin{equation}
\label{refine poison}
    {\theta c}^* = \arg\min_{\theta c}\sum^N_{i=1}L(f_{\theta c}(\alpha_{fi}), \beta_i) \quad \mathrm{s.t.}~~  (\alpha_{fi}, \beta_i) \in D_{fp}
\end{equation}
To generate mode mixture samples with control, we leverage the semi-discrete optimal transport (OT) algorithm to map a noise (Gaussian) distribution to latent code distribution and then locate latent code of mode mixture samples with angle filtering. An et al.~\cite{an2019ae} have suggested that the process of probability distribution transformation can be executed through semi-discrete OT mapping. The total cost of the transport mapping is defined below:
\begin{equation}
\label{transport cost}
T^* := \arg\min_{T_{\#\mu=\nu}}\int_{\Omega}c(x, T(x))d\mu(x)
\end{equation}

The source measure, $\mu$, is a Gaussian distribution defined on a convex domain, $\Omega$, which is a subset of $\mathbb{R}^d$. The target measure, $\nu$, is defined on a discrete set, $Y$, consisting of the points $y_1$, $y_2$, and so on up to $y_n$, each point being in $\mathbb{R}^d$. The source measure and the target measure have equal mass ($\mu(\Omega)=\sum_{i=1}^n\nu_i$). Based on Brenier's theorem, the Brenier potential function, $u_h(x)$ is given by $\max_{i=1}^{n} \{\pi_{h,i}(x)\}$ where $\pi_{h,i}(x)= \langle x, y_i \rangle + h_i$. It provides the semi-discrete OT map if the cost function is $c(x, y) = \frac{1}{2} \|x - y\|^2$. The unique optimizer $h$ can be optimized via gradient descent on a convex energy:
\begin{equation}
    \label{convex energy}
    E(h) = \int_0^h\sum_{i=1}^n\omega_i(\eta)d\eta_i-\sum_{i=1}^nh_i\nu_i
\end{equation}
where $\omega_i(\eta)$ is the $\mu-$volume of $W_i(\eta)$. $W_i(\eta)$ is formed by the projection of $u_h$ onto $\Omega$. The gradient of energy in Equation~\ref{convex energy} is approximated as $\nabla E(h) \approx (\omega_i(h)-\nu_i)^T$.

The semi-discrete OT maps each sample with a Gaussian noise distribution to a latent code distribution without generating any new samples. However, to accommodate for the generation of new samples, a piece-wise linear extension of the OT map is introduced. As stated earlier, the projection of the Brenier potential divides the given domain into multiple cells with samples in each cell mapped to the corresponding target $y_i$. We estimated the $\mu-$mass center $c_i$ by calculating the mean value of all samples inside $W_i$, and a point-wise mapping from $c_i$ to $y_i$ is constructed in this way. The vertices of polyhedrons surrounding $x$ are determined as the nearest $(d + 1)$ $\mu-$mass centers $c_{ik}, (k = 0, 1, .., d)$ in Euclidean distance, where $x$ is a random sample from a Gaussian noise distribution. Thus, the extended OT map $\bar{T}$ maps $x$ into $\sum_{k=0}^d\lambda_ky_{ik}$ where $\lambda_k = d^{-1}(x, c_{ik})/\sum_{k=0}^dd^{-1}(x, c_{ik})$. In particular, when different vertices of a polyhedron that surround $x$ belong to different modes, $y_i$ is the latent code of a mode mixture sample. To locate these latent codes, we calculated angles $\theta_{ik}$ between $\pi_{i0}$ and $\pi_{ik}$, $(k=1,2,...,d)$. If all angles $\theta_{ik}$ exceed a give threshold, the corresponding latent code belongs to a mode mixture sample. $\theta_{ik}$ is defined as:
\begin{equation}
    \theta_{ik} = \frac{\langle y_{i0}, y_{ik}\rangle}{\|y_{i0}\|\cdot\|y_{ik}\|}
\end{equation}

In this study, we set $d=1$ and manually adjust $\lambda_k$ to approximate the desired mode.

\paragraph{Poisoned Images Crafting:}
Crafting poisoned images involves adding imperceptible perturbations to clean images so that they approximate mode mixture samples in the latent space extracted by $\theta_e$. The transform function is described in Equation~\ref{transform function} and the optimization of perturbation is described in Equation~\ref{pert optim}
\begin{equation}
    \label{transform function}
    Tr(\alpha) = \alpha+\emph{pert}, \quad \|\emph{pert}\|_\infty\leq\epsilon
\end{equation}

\begin{equation}
    \label{pert optim}
    \emph{pert}^* = \arg\min_{\emph{pert}}\|\mathrm{feat}_{mm}-\theta_e(Tr(\alpha))\|^2
    \quad
    \mathrm{s.t.}~~  \|\emph{pert}\|_\infty\leq\epsilon
\end{equation}
where $\mathrm{feat}_{mm}$ stands for latent code of mode mixture samples. The ball with the L-infinity norm guarantees that poisoned images remain stealthy.  The transform function specified in Equation~\ref{transform function} formally makes our attack a perturbation-based backdoor approach. Optimizing $\emph{pert}$ is accomplished by employing the projected gradient descent (PGD) algorithm~\cite{madry2017towards}. Firstly, it optimizes the objective function in Equation~\ref{pert optim} using gradient descent, and secondly, projects the solution back into the $\epsilon-$neighborhood of the original image. To expedite training, we are inspired by Saha et al.~\cite{saha2020hidden}, and adopt a straightforward greedy algorithm that iterates over the mode mixture latent code and the perturbed clean image latent code, identifying the nearest pair for each other.


\section{Experiments}
\label{section4}
\subsection{Experimental Setup}
We performed experiments on four well-known datasets, namely Modified National Institute of Standards and Technology (MNIST)~\cite{lecun1998mnist}, Canadian Institute For Advanced Research (CIFAR-10)~\cite{krizhevsky2009learning}, German Traffic Sign Recognition Benchmark (GTSRB)~\cite{stallkamp2011german}, and Tiny ImageNet~\cite{le2015tiny}. To provide a detailed description of these datasets, we have presented their key features in Table~\ref{dataset discription}. The attack targets for this study are label 1 for the GTSRB dataset and label 0 for the other datasets.

Regarding the victim model, we chose ResNet-18~\cite{he2016deep}. We pre-trained the model using the stochastic gradient descent (SGD) optimizer with a $0.9$ momentum and $5e^{-4}$ weight decay. The initial learning rate was set to $0.1$, and it was reduced by $0.5$ when the outcome did not improve for $10$ epochs. The total number of epochs was 250. In the training of OT, computing the $\mu-$volume $\omega_i(h)$ of $W_i(h)$ was performed using the standard Monte Carlo method. The sample size was initially set at 20,000 and was increased when $E(h)$ ceased to decrease for 30 steps. In constructing $M_1$, $\lambda_k$ was set to $0.25$ for $y_{ik}$ from the attack target class and $0.75$ for $y_{ik}$ from other classes, and in constructing $M_2$, $\lambda_k$ was set to $0.5$. For the poisoned image generation, we set $\epsilon$ to $16$ (the pixel values range from 0 to 255) for Tiny ImageNet and $8$ to other datasets. We trained each batch of images for a maximum of $5,000$ iterations. The experiments were carried out on an Nvidia A100 GPU server.

 \begin{figure*}[t]
  \centering
  \includegraphics[width=0.8\textwidth]{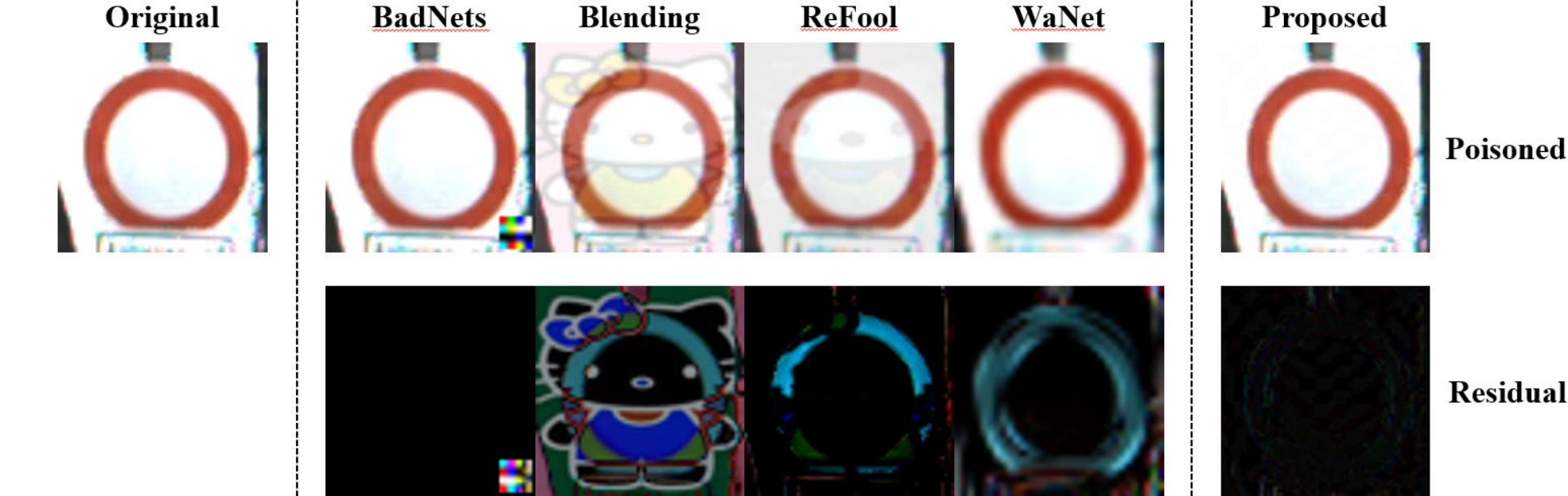}
  \caption{\label{stealthy compare} Visualization of poisoned images from different methods. From left to right: Original clean image, poisoned image by BadNets~\cite{gu2017badnets}, Blending~\cite{chen2017targeted}, ReFool~\cite{liu2020reflection}, WaNet~\cite{nguyen2021wanet}, and proposed method, respectively. The residual is amplified by $4\times$.}
\end{figure*}

\begin{figure*}[h]
  \centering
  \includegraphics[width=0.8\textwidth]{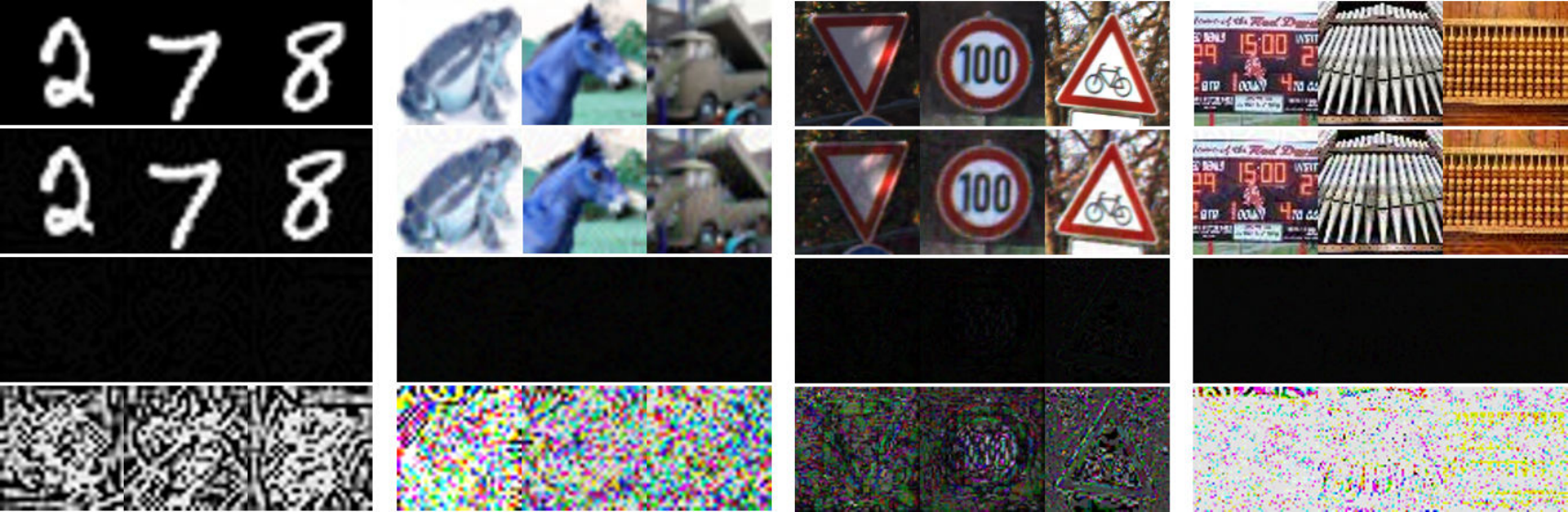}
  \caption{\label{stealthy alone} Visualization of poisoned images from our method. First row: original image samples. Second row: poisoned images. Third row: residual. Fourth row: normalized residual.}
\end{figure*}

\subsection{Performance of Attack}
To evaluate the performance of the attacks, we conducted tests to measure the clean accuracy and attack accuracy of various attack methods.
The comparison of the attack performance between our method and BadNets~\cite{gu2017badnets}, WaNet~\cite{nguyen2021wanet} and Data-Efficient attack~\cite{xia2022data} is shown in Tables~\ref{attack performance} and~\ref{attack performance2}.

The results in Table~\ref{attack performance} are obtained without any limitations on attackable parameters, while the results in Table~\ref{attack performance2} are obtained under the same number of attackable parameters. We keep $\theta e$ of the pre-trained clean classifier fixed, and only allow BadNets, WaNet, and Data-Efficient attack to optimize $
\theta c$. The results indicate that the proposed method achieves comparable clean accuracy and attack accuracy as the existing backdoor attack methods do while using significantly smaller amount of attackable parameters. In the scenario when users will only accept minor modifications on a trusted clean model for their safety, BadNets, WaNet, and Data-Efficient backdoor attacks performed poorly with low attack accuracy and dropped clean accuracy, while our method remains robust. A potential explanation is that previous methods for backdoor attacks depend on $\theta e$ to learn to cluster poisoned images in the latent space and then classify them using $\theta c$. In contrast, our approach directly modifies the latent space and implants the backdoor via mode mixture.

We noticed that the proposed method's attack accuracy on Tiny ImageNet declines slightly. We attribute this decline to the dataset's large number of classes and small number of samples per class, highlighting a limitation of our approach in such a scenario. To maintain stealthiness, we employed a method that injects backdoors by replacing original, clean features of the attack target class with mode mixture features. Since there are only 500 samples in one class of Tiny ImageNet, we can replace a maximum of 500 of these with mode mixture features. However, with 200 total classes, there is one target class and 199 non-target classes, and with 500 mode mixture features in total, the average number of mixture feature that mix target class and other classes is less than three (500/199$\approx$2.5). This poses great difficulties for optimizing poisoned images in testing phase with comparatively small perturbation budget. Although there is a limitation of performance on dataset that contains excessive large number of classes and small number if samples per class, our method perform significantly better under the same amount of attackable parameters, and also on stealthiness, as will be discussed later.

Table~\ref{attack performance3} clearly demonstrates the quantity of parameters utilized in embedding the backdoor. For each method, a pretrained clean model is initially employed, onto which the corresponding attack is subsequently applied. After the backdoor is successfully embedded into the network, parameters that fluctuated more than $1e-8$ was tallied. Table~\ref{attack performance3} illustrates that while other methods affected almost all parameters in the clean model, our approach only affected a negligible portion. Additionally, Table~\ref{attack performance4} presents the fluctuation range distribution of the altered parameters. The fluctuation range is quantified as $|\theta_{poisoned}-\theta_{clean}|\div|\theta_{clean}|\times100$, using $1e2$ as a milestone. Our method displays a larger segment of parameters with a fluctuation range below $1e2$. This implies that our method employs not only a smaller parameter count for backdoor embedding but also exhibits a reduced fluctuation range, while maintaining comparable attack accuracy. As a result, we can infer that under the proposed paradigm, our approach is stealthier.

\begin{table*}
  \centering
  \begin{tabular}{ccc}
    \hline
    \bf{Dataset} & \bf{With Mode Mixture} & \bf{W/O Mode Mixture} \\
    \hline
    CIFAR-10 & $1.28$ & $2.90$ \\
    GTSRB & $3.14$ & $5.77$ \\
    Tiny ImageNet & $6.03$ & $16.80$ \\
    \hline
  \end{tabular}
  \caption{\label{ablation1} $L_{2}$ norm of perturbation of poisoned images with and without mode mixture samples.}
\end{table*}

\begin{table*}
  \centering
  \begin{tabular}{cccccccc}
    \hline
    \multicolumn{1}{c}{\bf{Dataset}} & 
    \multicolumn{1}{c}{\bf{Model}} &
    \multicolumn{2}{c}{\bf{BadNets}} & 
    \multicolumn{2}{c}{\bf{WaNet}} &
    \multicolumn{2}{c}{\bf{Proposed}} \\ 
    \multicolumn{1}{c}{} & \multicolumn{1}{c}{} &
    \multicolumn{1}{c}{\bf{ARI}} & \multicolumn{1}{c}{\bf{Attack}} & 
                            \multicolumn{1}{c}{\bf{ARI}} & \multicolumn{1}{c}{\bf{Attack}} &
                            \multicolumn{1}{c}{\bf{ARI}} & \multicolumn{1}{c}{\bf{Attack}}\\
    \hline
    CIFAR-10 & ResNet18 & $0.74$ & $0.98$ & $0.99$ & $0.99$ & $0.48$ & $1.00$ \\
    Tiny ImageNet & ResNet18 & $1.00$ & $0.98$ & $1.00$ & $0.99$ & $0.40$ & $0.94$ \\
    \hline
  \end{tabular}
  \caption{\label{latent stealthy} Latent space stealthiness (ARI: Adjusted Rand Index)}
\end{table*}

\subsection{Stealthiness}
The stealthiness of the method in image space is facilitated by the proximity of mode mixture samples to the mode they blend, making it cost-efficient to optimize an image to approximate a sample that is nearby in latent space, as opposed to a distant one. Examples depicted in Figures~\ref{stealthy compare} and~\ref{stealthy alone} demonstrate the stealthiness of our approach. The poisoned images generated by our method have a higher visual similarity to clean images. Our method, besides being highly stealthy, boasts a high attack success rate on poisoned samples with very little amount of attackable parameters. More samples are shown in appendix.

\subsection{Ablation Study}
We conduct an ablation study to investigate the efficacy of using mode mixture samples in poisoned images crafting as opposed to normal samples. The fundamental rationale behind utilizing mode mixture samples is their close relationship with the modes they are mixing in latent space. When attempting an attack on an airplane image with the intention of misclassifying it as a dog, modifying its feature to resemble the mode mixture feature of an airplane and a dog in latent space proves to be more efficient. It requires lesser optimization compared to modifying its feature to directly imitate a dog, thus, conserving the perturbation budget. We evaluated the average $L_{2}$ norm of perturbations required to achieve successful attacks both with and without mode mixture samples. In the case of without mode mixture samples, all clean images' pixels are perturbed to approximate attack target class features in latent space rather than mode mixture features. The results of this study are presented in Table~\ref{ablation1}. From Table~\ref{ablation1} it is clear that with mode mixture samples, the perturbation are smaller, resulting in more stealthier poisoned images.

\subsection{Against Defense Methods}
We evaluate our approach using several defensive mechanisms: latent space defense, model mitigation defense, and sample detection defense.
\subsubsection{Latent Space Defense}
\begin{figure*}
  \centering
  \includegraphics[width=0.8\textwidth]{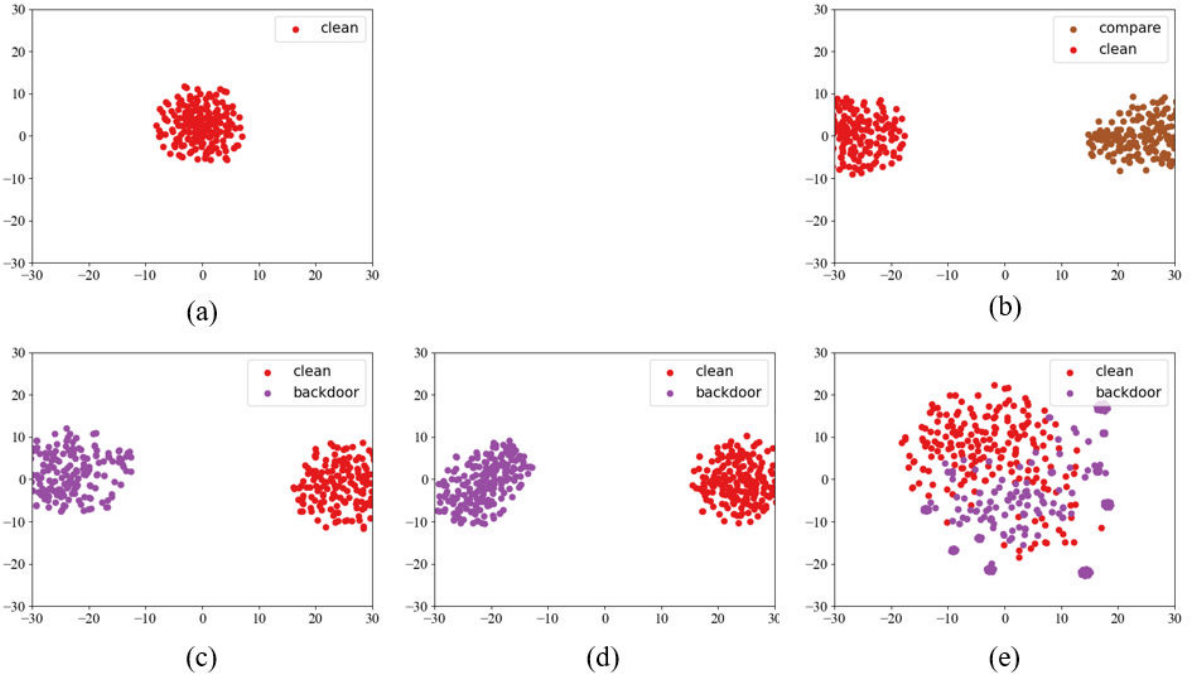}
  \caption{t-SNE embedding in latent space}
  \label{latent space}
\end{figure*}

Activation clustering, proposed by Chen et al.~\cite{chen2018detecting}, is a well-known defense method in the field of latent spaces. According to the authors, even though both clean and poisoned images are classified into the same class, they can still be distant and separable in the latent space. Therefore, they employed various dimensional reduction techniques on latent codes and applied K-Means to low-dimensional ones to distinguish between the two types of images. If the separation can be achieved with high accuracy, it indicates that the model has been backdoored. The stealthiness of our method in latent space is also facilitated by the proximity of mode mixture samples to the mode they blend. Crafting poisoned images involves optimizing their proximity to mode mixture samples around the target attack class in latent space, ensuring they are naturally located near the same range as the attack target class. In this study, we use the penultimate layer of the classifier as the latent space and utilize PCA and t-SNE for reducing dimensions, followed by K-Means with $k=2$ to cluster the data. 
Figure~\ref{latent space} illustrates the t-SNE embedding of clean and poisoned images in the latent space of tiny imagenet. Specifically, Figure~\ref{latent space}(a) shows the clean images belonging to the target class that cluster together in the latent space. In contrast, Figure~\ref{latent space}(b) exhibits clean images from two different classes, including the target class, which are widely scattered in the latent space. Figures~\ref{latent space}(c)-(e) present the t-SNE embedding of the clean and poisoned images in the latent space of BadNets, WaNet, and the proposed method, respectively. Notably, BadNets and WaNet visibly spread out within the same class label, resembling the distribution in Figure \ref{latent space}(b), making them easily distinguishable by the activation clustering method~\cite{chen2018detecting}. In contrast, the proposed method brings clean and poisoned images closer to each other in the latent space without clearly separating them, thanks to mode mixture latent modification. Consequently, the proposed method attains a better level of stealthiness compared to BadNets and WaNet in latent space.

\begin{figure}[t]
  \centering
  \includegraphics[width=0.4\textwidth]{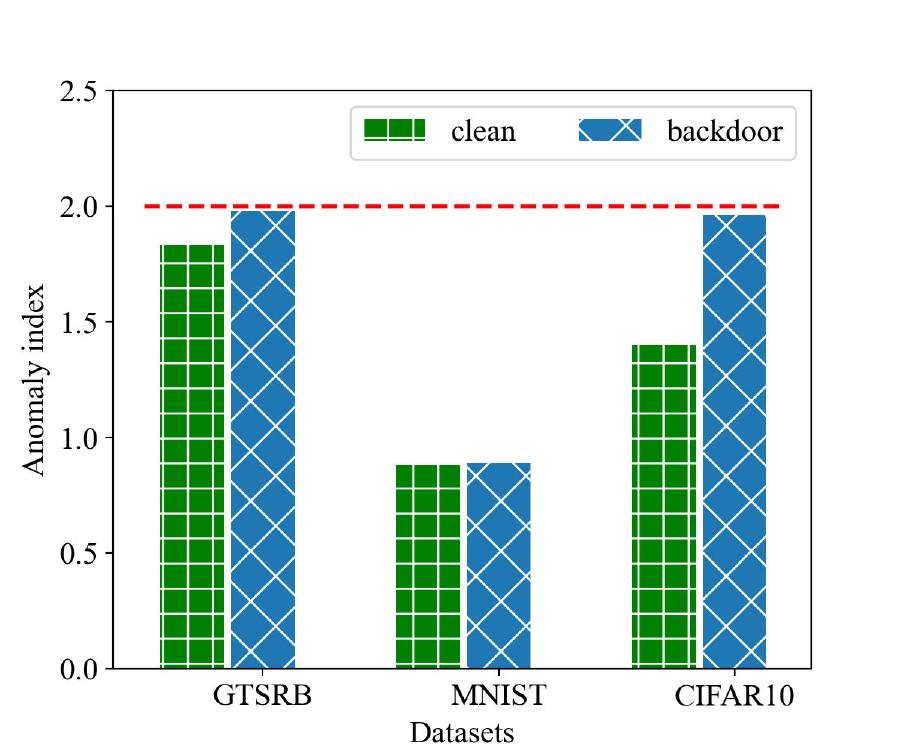}
  \caption{\label{neural} Defense experiments of proposed mode mixture latent modification against Neural Cleanse~\cite{wang2019neural}. As clean models, backdoor models do not exceed the predetermined anomaly index threshold.}
  
\end{figure}

For quantitative analysis, we adopt the Adjusted Rand Index (ARI), which measures the similarity between two sets of data, the cluster result and the ground truth label (clean or poisoned). It compares pairs of data points that are correctly or incorrectly classified in both sets. The score of ARI ranges from -1 to 1, with a score of 1 indicating perfect agreement and a score of 0 indicating random agreement. ARI scores closer to 0 are preferred in resisting latent space defense. Therefore, it is concluded that our proposed method is better equipped to handle defenses in the latent space while maintain high attack accuracy compared with BadNets and WaNet, as shown in Table~\ref{latent stealthy}.

While our stealthiness in latent space may not rival that of previous works that focus on latent stealthiness, such as Adversarial Embedding~\cite{shokri2020bypassing} and Latent Attack~\cite{doan2021backdoor}, we contend that stealthiness in latent space is not the primary focus of this study. However, their stealthiness in latent space is achieved with a significantly larger budget of attackable parameters. In contrast, our method remains stealthy in latent space while utilizing fewer attackable parameters.

\begin{figure*}
  \centering
  \includegraphics[width=0.8\textwidth]{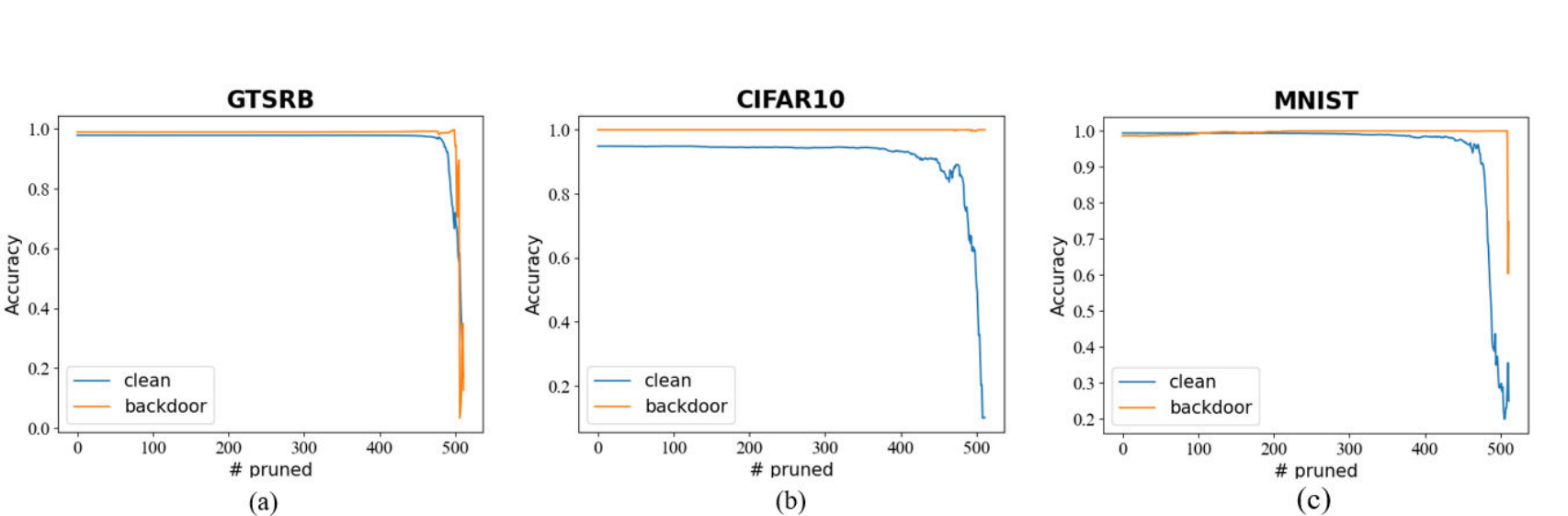}
  \caption{\label{fineprune} Fine-pruning results on GTSRB, CIFAR-10, and MNIST}
\end{figure*}

\begin{figure*}
  \centering
  \includegraphics[width=0.8\textwidth]{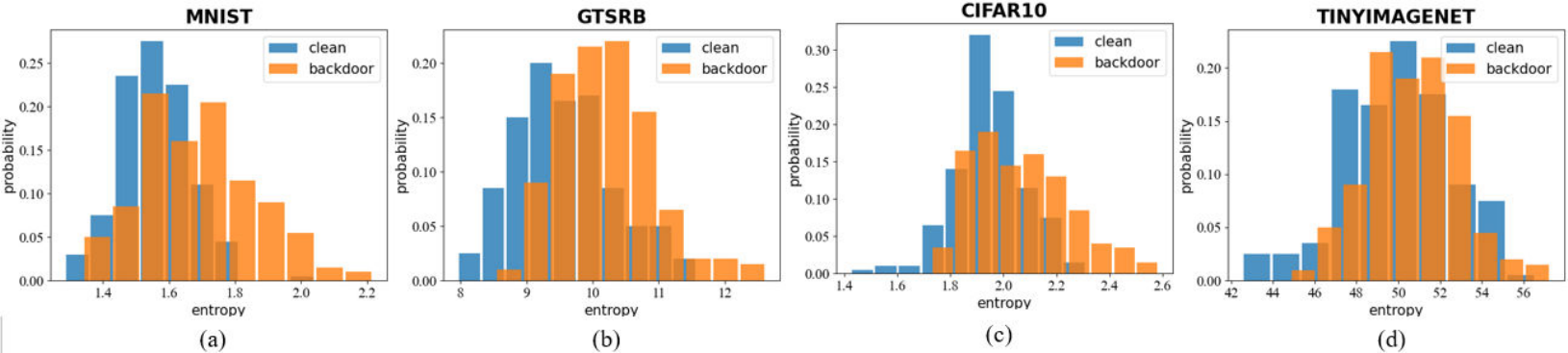}
  \caption{\label{STRIP}STRIP results on MNIST, GTSRB, CIFAR-10 and TinyImageNet}
\end{figure*}

\begin{figure}
  \centering
  \includegraphics[width=0.45\textwidth]{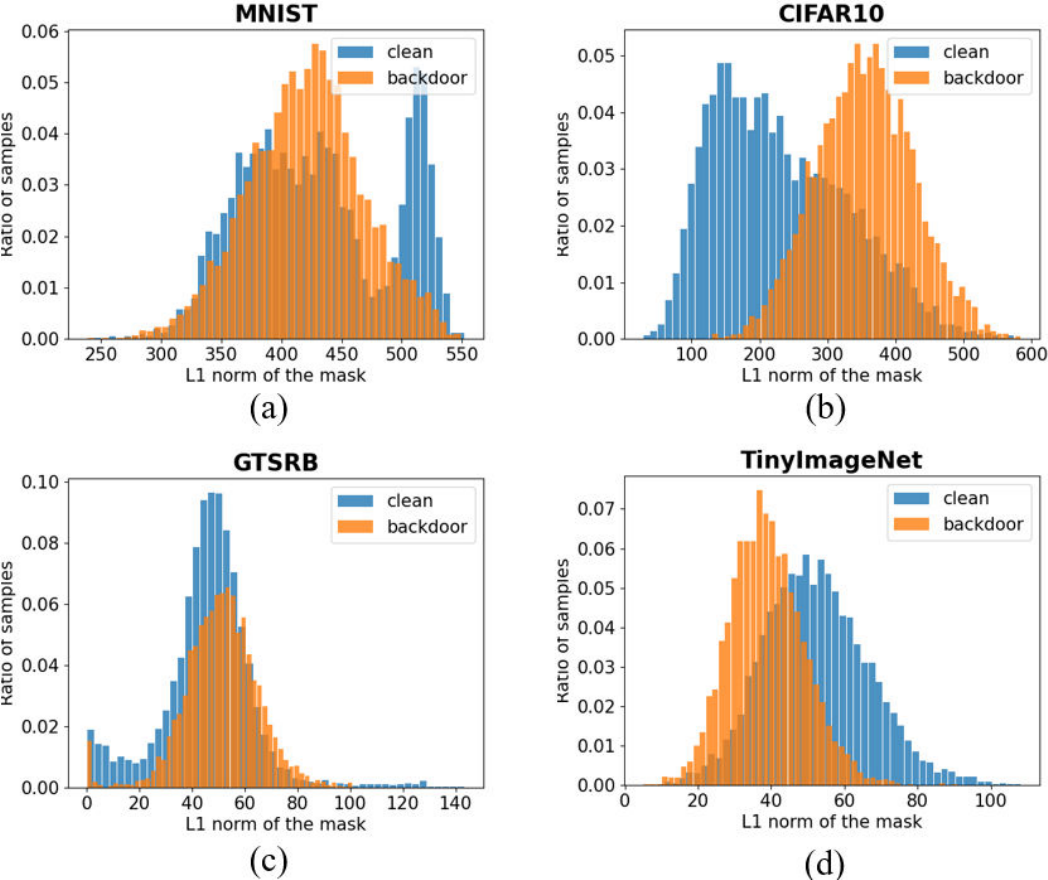}
  \caption{\label{CD Defense} Distributions of the $L_1$ norm of $m$ of clean and poisoned samples.}
\end{figure}

\subsection{Model Mitigation Defense}
Neural cleanse proposed by Wang et al~\cite{wang2019neural}. is a popular model mitigation defense. The model quantifies the smallest perturbation required to manipulate the classification results of all images in a dataset toward a specific target label. Subsequently, an anomaly index quantifies the model's anomaly degree. If the anomaly index of a model surpasses the predefined threshold 2, there is a $95\%$ chance of it being an outlier. 

The figures displayed in Figure~\ref{neural} demonstrate that the anomaly index of both the clean and backdoor models in GTSRB and MNIST are almost identical to each other. While the anomaly index of the backdoor model in CIFAR10 is higher than the clean model, it is still within the specified threshold. Thus, it is evident from Figure~\ref{neural} that our proposed method is impervious to neural cleansing.

Fine-pruning, proposed by Liu et al.~\cite{liu2018fine}, is another well-known model mitigation defense method. Given a model, fine-pruning first prunes the neurons that remain inactive for clean inputs, and then applies fine-tuning to the trimmed network. Eliminating dormant neurons can eliminate extra neurons that are not necessary for predicting clean inputs but may be indispensable for predicting poisoned inputs. In addition, if neurons are essential for both clean and poisoned inputs, fine-tuning can also remove the characteristics specific to poisoned inputs. Figure~\ref{fineprune} (a-c) displays the fine-pruning results on GTSRB, CIFAR-10, and MNIST, respectively. We pruned the last convolutional layer of the backdoor model as in previous studies~\cite{nguyen2021wanet, doan2021lira}. The backdoor models trained on CIFAR-10 and MNIST contained 512 neurons, whereas for GTSRB, 2048 neurons needed to be pruned. Channel-wise pruning is employed, which results in the pruning of 1 neuron and 4 neurons in CIFAR-10/MNIST and GTSRB each time, respectively.

\subsubsection{Sample Detection Defense}

STRong Intentional Perturbation (STRIP), proposed by Gao et al.~\cite{gao2019strip}, is a well-known defense method in the field of sample detection defense. The test image is perturbed using a set of clean images randomly selected from the dataset. In the case of benign images, the perturbation produces unpredictability in the prediction outcomes, which results in high normalized entropy. However, for backdoor-attacked images, the prediction results are more likely consistent due to the presence of a backdoor trigger, leading to a low normalized entropy in the prediction outputs. Figure~\ref{STRIP}(a-d) displays the STRIP results on MNIST, GTSRB, CIFAR-10, and TinyImageNet, respectively.

Figure~\ref{STRIP} demonstrates that the poisoned images in the proposed attack method cover the similar range of normalized entropy as clean images. Based on the results, we can conclude that our method is resistant against STRIP.

Huang et al.~\cite{huang2023distilling} introduced Cognitive Distillation (CD) as a method to extract a minimal pattern from an input image determining the model's output. A mask $m$ is employed to derive a minimal pattern known as Cognitive Pattern (CP). They discovered that the CPs of backdoor samples tend to be smaller than those of clean samples. Therefore, they propose a defense mechanism for poisoned sample detection by computing the $L_1$ norm of $m$. The distributions of the 
$L_1$ norm of $m$ of clean and poisoned samples should be easily distinguishable, with the $L_1$ norm of $m$ for poisoned samples expected to be smaller than that for clean samples. Figure~\ref{CD Defense}(a-d) displays the distributions of $L_1$ norm of $m$ for clean (in blue) and poisoned (in orange) samples generated by our method.

Figure~\ref{CD Defense} clearly shows that the distributions of the $L_1$ norm of $m$ of clean and poisoned images, generated by our method, covers a similar range and is not separable. Even though in the case of CIFAR10, the distributions might appear to be separable, $L_1$ norms of the backdoor samples' masks are larger than those of the clean samples, which may cause incorrect predictions using this defense method.

\section{Conclusions}
\label{section6}
This study introduce an insidious backdoor attack paradigm that requires very little attackable parameters and propose a novel backdoor attack method with mode mixture latent modification under such paradigm. Additionally, we utilize semi-discrete OT and its piece-wise linear extension to map a Gaussian noise distribution to latent code distribution and thus generate mode mixture samples with control. We also highlight the potential application of mode mixture phenomenon, despite its avoidance as a mainstream practice. Our backdoor model exhibited high accuracies in both clean and attack scenarios with limited attackable parameters, and demonstrated resilience against commonly deployed defense mechanisms, as observed in experiments conducted on four benchmark datasets.

Despite our study being exclusively conducted on the image classification task, it is noteworthy that our method is readily adaptable to other domains, including facial and speech recognition. This flexibility arises from the manipulation occurring in the latent space, making it independent of specific network architectures.

Our work has several limitations. First of all, for datasets with a large number of classes and a small number of samples per class, the efficacy of our approach decreased. The average number of mode mixture features mixing attack target class and other classes will be insufficient for the optimization of poisoned samples. Consequently, it affects the stealthiness of poisoned samples and the success rate of the attack. Further more, our method focuses on all-to-one attack setting under the proposed attack paradigm, but does not explore all-to-all attack setting. These limitations will be addressed in our future studies.





\bibliographystyle{plain}
\bibliography{usenix}

\section*{Appendix}
\section*{A. Dataset and Models}
\subsection*{MNIST}
The Modified National Institute of Standards and Technology (MNIST)~\cite{lecun1998mnist} database is an expansive source of handwritten digits, frequently utilized for the training of numerous image processing systems. It comprises 60,000 training images, in addition to 10,000 testing images. Each image signifies a digit in the range of 0 to 9 and is represented by a $28\times28$ grayscale grid. The MNIST dataset holds a popular position for benchmarking classification algorithms, primarily within the realms of machine learning and computer vision.

\subsection*{CIFAR-10}
The Canadian Institute for Advanced Research-10 (CIFAR-10)~\cite{krizhevsky2009learning} serves as a prevalent resource for machine learning and computer vision applications. It encapsulates 60,000 colour images, each with 32x32 pixels, distributed across 10 distinct classes, each comprising 6,000 images. These classes encompass airplanes, cars, birds, cats, deer, dogs, frogs, horses, ships, and trucks respectively. Divided into 50,000 training images and 10,000 testing ones, the CIFAR-10 dataset is routinely employed to train machine learning and computer vision models, as well as to benchmark emerging algorithms.

\subsection*{GTSRB}
The German Traffic Sign Recognition Benchmark (GTSRB)~\cite{stallkamp2011german}, a multi-class, single-image classification challenge, was showcased at the International Joint Conference on Neural Networks (IJCNN) 2011. This dataset encompasses more than 50,000 images delineating 43 unique classes of traffic signs including, but not restricted to, speed limits, prohibitory signs, mandatory signs, and danger signs. Each image, varying in size and subjected to different lighting conditions and environments, presents a distinct challenge for the creation and validation of traffic sign recognition algorithms. The objective is to construct a model capable of accurately categorizing these images into their corresponding classes, making the GTSRB an extensively employed benchmark within machine learning and computer vision fields.

\subsection*{TinyImageNet}
The Tiny ImageNet dataset~\cite{le2015tiny}, frequently utilized for benchmarking machine learning algorithms, provides a compact variant of the larger ImageNet dataset. Predominantly, the ImageNet dataset finds its extensive application in machine learning, especially in image recognition tasks.
Tiny ImageNet comprises 200 distinct classes, with each containing 500 training images, 50 validation images, and 50 test images, aggregating to a total of 120,000 images. All images are color-based and uphold a resolution of 64x64.
The reduced size of Tiny ImageNet, relative to the original ImageNet, expedites testing and iteration processes in machine learning models. Consequently, it is a popular choice for projects and research necessitating more efficiently manageable image classification tasks.

\subsection*{Pre-trained Models}

\begin{table}[h]
  \centering
  \begin{tabular}{cccc}
    \hline
    \multicolumn{1}{c}{\bf{Model}} &
    \multicolumn{1}{c}{\bf{Dataset}} & 
    \multicolumn{1}{c}{\bf{\#Parmeters}} & 
    \multicolumn{1}{c}{\bf{Input Size}}\\
    \hline
    Resnet18 & MNIST & $11.2M$ & $28\times28\times3$\\
    Resnet18 & CIFAR-10 & $11.2M$ & $32\times32\times3$\\
    Resnet18 & GTSRB & $11.5M$ & $64\times64\times3$\\
    Resnet18 & TinyImageNet & $11.5M$ & $64\times64\times3$\\
    \hline
  \end{tabular}
  \caption{\label{pre-trained} Pre-trained models}
\end{table}

\section*{B. Poisoned Images}
\begin{figure}[h]
  \centering
  \includegraphics[width=0.45\textwidth]{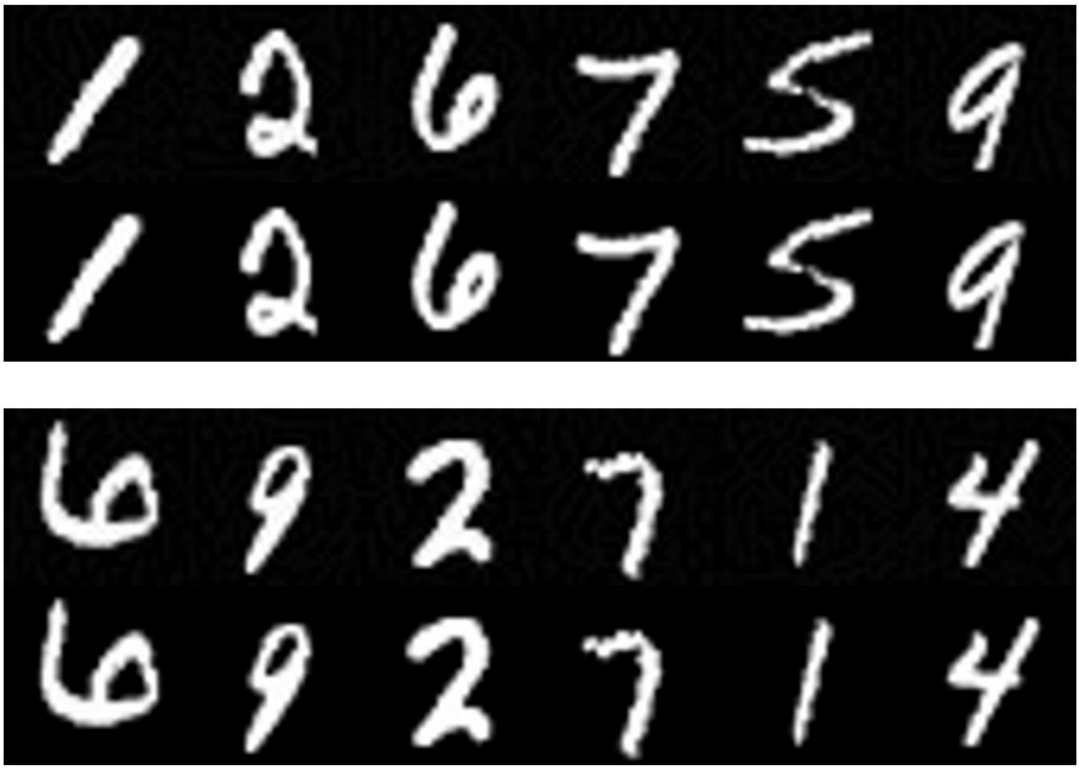}
  \caption{More poisoned image samples from MNIST. Poisoned images in row 1 and original images in row2. All poisoned images are classified as 0.}
\end{figure}

\begin{figure}[h]
  \centering
  \includegraphics[width=0.45\textwidth]{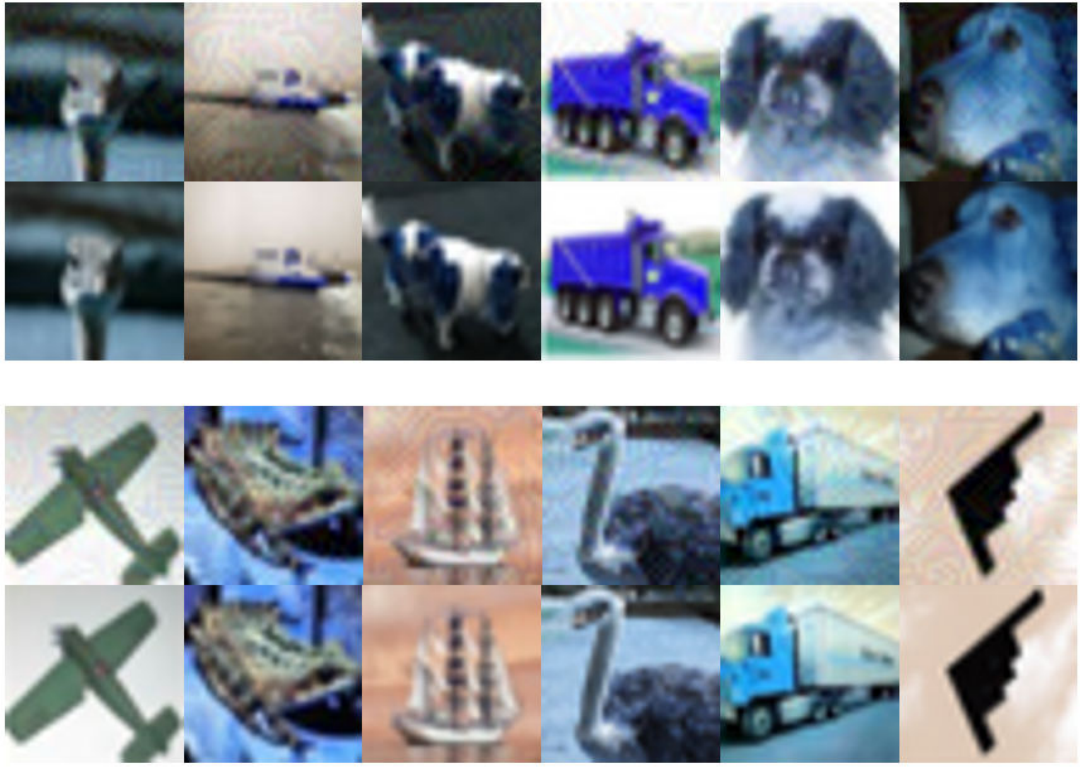}
  \caption{More poisoned image samples from CIFAR-10. Poisoned images in row 1 and original images in row2. All poisoned images are classified as airplane (label 0).}
\end{figure}

\begin{figure}[h]
  \centering
  \includegraphics[width=0.45\textwidth]{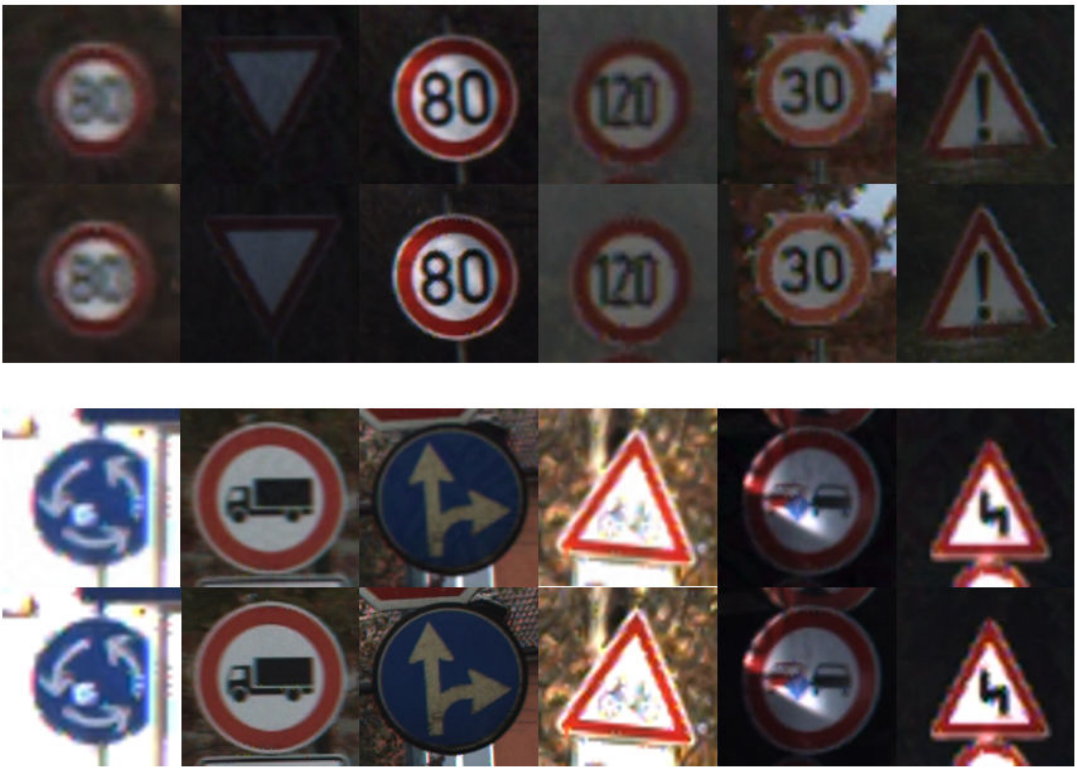}
  \caption{More poisoned image samples from GTSRB. Poisoned images in row 1 and original images in row2. All poisoned images are classified as speed limit 30 (label 1).}
\end{figure}

\begin{figure}[t]
  \centering
  \includegraphics[width=0.45\textwidth]{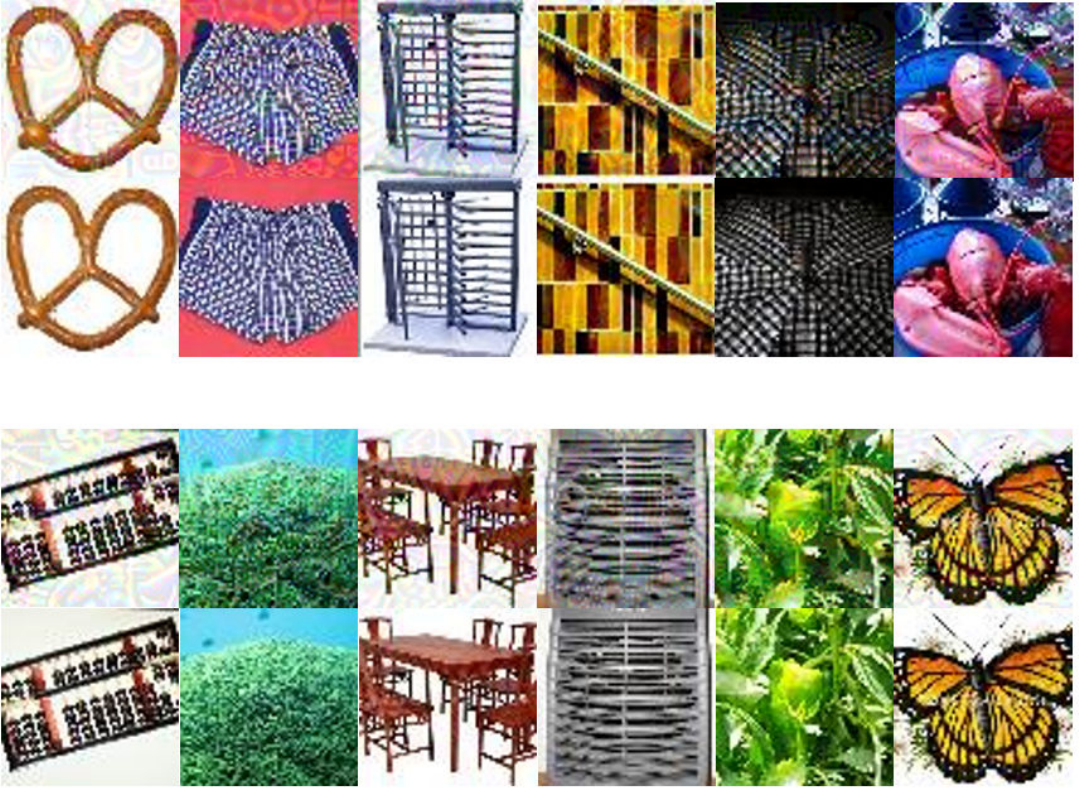}
  \caption{More poisoned image samples from TinyImageNet. Poisoned images in row 1 and original images in row2. All poisoned images are classified as label 0.}
\end{figure}

\end{document}